# Transport Through a Quantum Dot with Electron-Phonon Interaction

Levente Máthé[a], Ioan Grosu[a,*]

[a]*Department of Physics, "Babeş-Bolyai" University, str. M. Kogălniceanu 1, Cluj-Napoca 400084, Romania*

**Abstract**

We theoretically study the electrical transport properties of a single level quantum dot connected to two normal conducting leads, which is coupled to the lattice vibrations. We determine the current through the quantum dot in two different situations: time-independent and time-averaged. In all situations we consider three cases: when there is no electron-phonon interaction, when the dot electrons interact with optical phonons or when they interact with acoustic phonons. At finite temperatures we take into account the temperature dependence of the chemical potential. We treat the electron-phonon interaction by the canonical transformation method. In the case of electron-longitudinal optical phonon interaction the spectrum contains a subpeak. In the case of electron-acoustic phonon interaction the spectrum is continuous. In the time-averaged situation many parasite peaks appear in the spectrum, due to the external time-modulation.

## 1. Introduction

Today's manufacturing techniques allow making artificial structures of nanometer sizes with many interesting transport properties and many applications in modern electronics [1,2]. Quantum dots were discovered by A. Ekimov and A.A. Onushchenko in glass crystals [2]. For these nanostructures exists a quantum confinement for motions of particles, charge carriers, in all three spatial directions. This confinement can be caused by the presence of an interface between different semiconductor materials, an electrostatic potential applied by contact electrodes or external strains [3]. It was shown that the electronic and optical properties of quantum dots strongly depend on their size and shape and that the tunnelling effects arise through the barriers [4].

An important problem in the study of these structures is the effect of electron-phonon scattering on the tunnelling current [1], because these phonon modes (lattice vibrations) can strongly influence the electrical properties of such a system. The electron-phonon interaction in the quantum dots has been studied by many groups [1,5-10, 22], mostly just for optical phonons in stationary cases. A few groups have investigated interaction of electrons with acoustic phonons [10-14]. Generally, the electron-acoustic coupling has two contributions: deformation potential coupling and piezo-electric coupling [15]. Most of the study was focused only on the piezo-electric coupling, because most of the semiconducting materials exhibit piezoelectricity [16]. Unfortunately, the literature on time-dependent non-equilibrium transport treated with Green functions is very restricted unlike in stationary cases [17-21]. The majority of these works use canonical transformation method for the treatment of electron-phonon interactions [7-10].

This work is focused on the description of time-independent and time-averaged electron transport through a quantum dot with optical and acoustic phonon interactions.

* Corresponding author. Tel.: +4-026-440-5300; fax: +4-026-459-1906.
  *E-mail address:* ioan.grosu@phys.ubbcluj.ro

## 2. Theoretical model

*2.1. Hamiltonian*

We consider a single level quantum dot, which is connected to two normal conducting leads, left (*L*) and right (*R*) lead, respectively. The dot is also coupled to the lattice vibrations, thus the electrons can interact with these phonon modes. The schematic representation of our considered system is presented in Fig. 1. We are interested in the determination of time-independent and time-averaged currents flowing from the non-interacting leads to the interacting quantum dot. For this we start our calculations with the suppositions made by A.-P. Jahuo et al. [17].

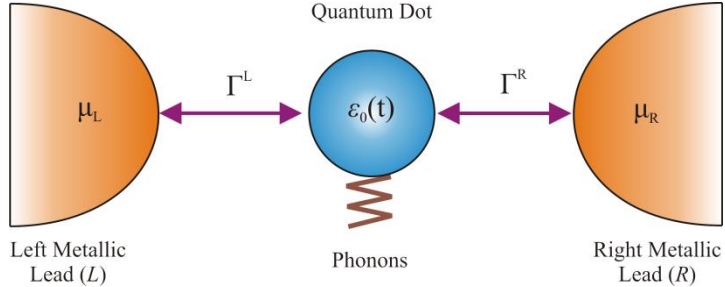

Fig. 1. The schematic diagram of the considered system. The single level quantum dot is connected to two metallic contacts with couplings $\Gamma^L$ and $\Gamma^R$, and the dot electrons interact with the phonon modes.

We set the bias voltage between the leads as: $\mu_L = \mu(T) + eV$ and $\mu_R = \mu(T)$, where $\mu_L$ ($\mu_R$) is the chemical potential in contact *L* (*R*), and we set the temperatures in the leads to be equal, so $T_L = T_R = T$. In following calculations, we consider the temperature dependence of the chemical potential [23]. Next, we omit the spin indices, and neglect the electron-electron interaction, thus the total Hamiltonian for the system can be written as [17]:

$$H = \sum_{\mathbf{k},\alpha \in L,R} \varepsilon_{\mathbf{k}\alpha}(t) c^+_{\mathbf{k}\alpha} c_{\mathbf{k}\alpha} + \sum_{\mathbf{k},\alpha \in L,R} \left[ V_{\mathbf{k}\alpha}(t) c^+_{\mathbf{k}\alpha} d + V^*_{\mathbf{k}\alpha}(t) d^+ c_{\mathbf{k}\alpha} \right] + \\ + \varepsilon_0(t) d^+ d + d^+ d \sum_{\mathbf{q}} M_{\mathbf{q}} \left( a_{\mathbf{q}} + a^+_{-\mathbf{q}} \right) + \sum_{\mathbf{q}} \hbar \omega_{\mathbf{q}} a^+_{\mathbf{q}} a_{\mathbf{q}} \qquad (1)$$

The first term represents the contact Hamiltonian, $H_C$, which describes the electrons in the left and right non-interacting metallic leads, where $c_{\mathbf{k}\alpha}^+$ and $c_{\mathbf{k}\alpha}$ denote the creation and annihilation operators of electrons, *k* is the wave vector (momentum) of the electrons and $\alpha$ labels the channel in the left (*L*) and right (*R*) metallic lead. The $\varepsilon_{\mathbf{k}\alpha}(t)$ are the single-particle energies of conduction electrons in the leads, and they are time-dependent [17]:

$$\varepsilon_{\mathbf{k}\alpha}(t) = \varepsilon^0_{\mathbf{k}\alpha} + \Delta_{\mathbf{k}\alpha}(t) \qquad (2)$$

where $\Delta_{\mathbf{k}\alpha}(t)$ is the external time modulation.

The second term in the total Hamiltonian is the tunneling Hamiltonian, $H_T$, and describes the coupling of the central region (quantum dot) to the contacts. Here $V_{\mathbf{k}\alpha}(t)$ represents the time-dependent tunneling matrix elements. The $d^+$ and $d$ represent the creation and annihilation operators of quantum dot electrons. The third term in the expression of the total Hamiltonian models the central region, $H_{cen}$. The fourth term describes the interaction of an electron with phonons, $H_{EP}$. $M_{\mathbf{q}} \equiv M(\mathbf{q})$ is the electron-phonon coupling constant. We note that the electron-phonon interaction is negligible, when $|\mathbf{q}| > 2\pi/L$, where *L* is the size of quantum dot. The last term represents the free-phonon contribution, $H_{ph}$. Here $\omega_{\mathbf{q}}$ is the frequency of the phonon modes. (In the following $\hbar=1$).

*2.2. Current Formulas*

The general time-dependent expression for the current which flows from the left contact is [17-19, 24]:

$$J_L(t) = -\frac{2e}{h} \int_{-\infty}^{t} dt' \int d\varepsilon \, \text{Im} Tr \left\{ e^{-i\varepsilon(t'-t)} \mathbf{\Gamma}^L(\varepsilon,t',t) \cdot \left[ \mathbf{G}^r(t,t') \cdot f_L(\varepsilon) + \mathbf{G}^<(t,t') \right] \right\} \quad (3)$$

where $\mathbf{G}^r(t,t')$ is the central region Green's function, and $\mathbf{G}^<(t,t')$ represents the lesser Green's function. $f_L(\varepsilon)$ is the Fermi function in left lead, and $\mathbf{\Gamma}^L(\varepsilon,t',t)$ is the level-width function for the left side, which depends on the tunnel matrix elements and gives the strength of the tunnel coupling between the left metallic contact and the quantum dot:

$$\mathbf{\Gamma}^L(\varepsilon,t',t) = 2\pi \sum_{\alpha \in L} \rho_\alpha(\varepsilon) V_{\alpha,n}(\varepsilon,t) V_{\alpha,m}^*(\varepsilon,t') \exp\left[ -i\int_t^{t'} dt_1 \Delta_\alpha(\varepsilon,t_1) \right] \quad (4)$$

where $\Delta_{\mathbf{k}\alpha}(t) \rightarrow \Delta_\alpha(\varepsilon,t)$ is the external time modulation in a given lead and $\rho_\alpha(\varepsilon)$ is the density of states in the leads and $V_{\mathbf{k}\alpha,n}(t) \equiv u(t) \cdot V_{\alpha,n}(\varepsilon_\mathbf{k})$ and we will assume that $u(t) = 1$. The level-width function for right side can be defined in the similar way, by replacing the indices $L$ with $R$. We note that the boldface notation of the level-width function and Green functions indicates that they are matrices. We can write an analogous formula for the current, $J_R(t)$, which flows from the right lead through the right barrier into the quantum dot. The time-independent current is:

$$J = -\frac{2e}{h} \frac{\Gamma^L \cdot \Gamma^R}{\Gamma^L + \Gamma^R} \int d\varepsilon \left[ f_L(\varepsilon) - f_R(\varepsilon) \right] \cdot \text{Im} G^r(\varepsilon) \quad (5)$$

The only task is the determination of the retarded Green function. The time-averaged current is given by:

$$\langle J_L(t) \rangle = -\frac{2e}{h} \frac{\Gamma^L \cdot \Gamma^R}{\Gamma^L + \Gamma^R} \int d\varepsilon \left[ f_L(\varepsilon) - f_R(\varepsilon) \right] \times \text{Im} \langle A(\varepsilon,t) \rangle \quad (6)$$

where:

$$A(\varepsilon,t) \equiv \int_{-\infty}^{t} dt' G^r(t,t') \exp\left[ i\varepsilon(t-t') + i\int_{t'}^{t} dt_1 \Delta(t_1) \right] \quad (7)$$

*2.3. Treatment of Electron-Phonon Interaction*

We eliminate the electron-phonon interaction in the total Hamiltonian by applying a canonical transformation method [7-10,25-26], which leads to separate the Green function into an electron and phonon part, respectively [16]. In this method the electron-phonon coupling can be strong, unlike in the perturbation methods [21].

To decouple the electron-phonon interaction we define a new transformed Hamiltonian $\bar{H} = e^S \cdot H \cdot e^{-S}$. For the S operator we have [16]:

$$S = d^+ d \sum_\mathbf{q} \left( M_\mathbf{q} / \omega_\mathbf{q} \right) \left[ a_{-\mathbf{q}}^+ - a_\mathbf{q} \right] \quad (8)$$

Thus, the new transformed Hamiltonian, $\bar{H}$, can be divided into an electron ($\bar{H}_{el}$) and phonon ($\bar{H}_{ph}$) part as $\bar{H} = \bar{H}_{el} + \bar{H}_{ph}$. The electron part of the transformed Hamiltonian is equal to $\bar{H}_{el} \approx H_C + \bar{H}_{cen} + H_T$, where the new transformed central region Hamiltonian is $\bar{H}_{cen} = \varepsilon_{EP}(t) d^+ d$, with the new single-level dot energy:

$$\begin{cases} \varepsilon_{EP}(t) = \varepsilon_0(t) - \Delta_{ph} \\ \Delta_{ph} \equiv \sum_{\mathbf{q}} \left( M_{\mathbf{q}}^2 / \omega_{\mathbf{q}} \right) \end{cases} \quad (9)$$

where $\Delta_{ph}$ is the energy shift. The phonon part remains unchanged, so $\bar{H}_{ph} = H_{ph}$.

By ignoring the Fermi sea, for the retarded Green function which is decoupled into an electron ($G^{(r)el}(t,t')$) and phonon ($G^{ph}(t,t')$) part, we can write [7,10,17-18]:

$$G^r(t,t') = -i\theta(t-t') \left\langle \left\{ d_{el}(t), d_{el}^+(t') \right\} \right\rangle_{el} \left\langle \left\{ X_{ph}(t) \cdot X_{ph}^+(t') \right\} \right\rangle_{ph} \equiv G^{(r)el}(t,t') \cdot G^{ph}(t,t') \quad (10)$$

where the new operators are $d_{el}(t) = \exp(i\bar{H}_{el} t) \cdot d \cdot \exp(-i\bar{H}_{el} t)$ and $X_{ph}(t) = \exp(i\bar{H}_{ph} t) \cdot X \cdot \exp(-i\bar{H}_{ph} t)$ with $X = \exp[-\sum_{\mathbf{q}} (M_{\mathbf{q}}/\omega_{\mathbf{q}})(a_{-\mathbf{q}}^+ - a_{\mathbf{q}})]$. In the wide-band limit the electron part of retarded Green function can be written as [8,10,17-18]:

$$G^{(r)el}(t,t') = -i\theta(t-t') \exp\left[-i \int_{t'}^{t} dt_1 \, \varepsilon_{EP}(t_1) \right] \exp\left[ -\frac{1}{2} \left( \Gamma^L + \Gamma^R \right) (t-t') \right] \quad (11)$$

The phonon part of retarded Green function is evaluated as [7,10,16,21]:

$$G^{ph}(t,t') = \left\langle \left\{ X_{ph}(t) \cdot X_{ph}^+(t') \right\} \right\rangle_{ph} = \exp[-\phi(t-t')] \quad (12)$$

$$\phi(t-t') = \sum_{\mathbf{q}} \left( \frac{M_{\mathbf{q}}}{\omega_{\mathbf{q}}} \right)^2 \left\{ N_{\mathbf{q}} \left[ 1 - \exp(i\omega_{\mathbf{q}}(t-t')) \right] + (N_{\mathbf{q}} + 1) \left[ 1 - \exp(-i\omega_{\mathbf{q}}(t-t')) \right] \right\} \quad (13)$$

is a time-dependent function, with $N_{\mathbf{q}}$ the average number of phonons.

*2.3.1. Interaction with longitudinal optical phonons*

The optical phonons are polarization waves inside an ionic crystal, thus they correspond to the vibrations of oppositely charged ions [26].

The electron-phonon coupling constant is given by the relation [16,26-27]:

$$M(\mathbf{q}) = \frac{i}{q} \left\{ \frac{e^2 \omega_0}{2\Omega \varepsilon_0} \left[ \frac{1}{\varepsilon_r(\infty)} - \frac{1}{\varepsilon_r(0)} \right] \right\}^{1/2} \quad (14)$$

where $\varepsilon_0$ is the dielectric constant of vacuum, $\varepsilon_r(\infty)$ and $\varepsilon_r(0)$ are the relative dielectric constants at high and law frequencies, $\Omega$ is the volume of the crystal, and $\omega_0$ is the constant angular frequency.

*2.3.2. Interaction with acoustic phonons*

The dispersion relation of acoustic phonons can be approximated as $\omega(\mathbf{q}) \approx c \cdot |\mathbf{q}|$ where $c$ is the speed of sound in the medium. The squared modulus of electron-phonon coupling constant can be expressed as [10,12,14]:

$$|M(\mathbf{q})|^2 = g\sqrt{\frac{2}{\pi}} \frac{1}{L^2} \frac{\pi^2 c^2}{\Omega |\mathbf{q}|} \frac{1}{|\mathbf{q}|^2 + (1/L)^2} \tag{15}$$

Here $g$ is a dimensionless coupling constant and depends on the crystal structure, and $\Omega$ is the volume of the crystal.

## 3. Calculations and Results

In this section, we present our results for the transport properties of quantum dot with electron-phonon interaction. We note that in order to simplify the numerical calculations we fix the level-width functions of contacts at equal value, so $\Gamma^L = \Gamma^R = \Gamma/2$ and we measure all energy quantities in unity of $\Gamma = \Gamma^L + \Gamma^R$. It is important to note that for all finite temperature cases the chemical potential depends on the temperature and we fix $\varepsilon_F = 1\,\Gamma$.

*3.1. Time-independent current*

3.1.1. Interaction with longitudinal optical phonons

First, we calculate some important quantities, like the energy shift. We convert the summation over $\mathbf{q}$ into an integral and we obtain:

$$\Delta_{ph} \equiv \sum_{\mathbf{q}} \frac{|M_\mathbf{q}|^2}{\omega_0} = \frac{e^2}{2\pi L \varepsilon_0}\left(\frac{1}{\varepsilon_r(\infty)} - \frac{1}{\varepsilon_r(0)}\right) \tag{16}$$

which is a constant for a given material. The coupling constant is defined as:

$$g \equiv \sum_{\mathbf{q}} \left|\frac{M_\mathbf{q}}{\omega_0}\right|^2 = \frac{\Delta_{ph}}{\omega_0} \tag{17}$$

From these relations we can define another parameter [10], the strength of coupling ($\beta$):

$$\beta^2 \equiv \sum_{\mathbf{q}} |M_\mathbf{q}|^2 = \omega_0 \Delta_{ph} \tag{18}$$

It is important to note that the strength of coupling is a phonon frequency-dependent quantity. Using the introduced parameters, we obtain [7,10,16], at zero temperature:

$$G^r(\varepsilon) = e^{-g} \sum_{n=0}^{\infty} \frac{g^n}{n!} \frac{1}{\left[\varepsilon - (\varepsilon_0 - \Delta_{ph}) - \omega_0 \cdot n + i\Gamma/2\right]} \tag{19}$$

By substituting Eq. (19) in Eq. (5) the conductance can be expressed. At finite temperatures, the retarded Green function can be expressed as:

$$G^r(\varepsilon) = e^{-g(2N_0+1)} \sum_{p=-\infty}^{\infty} \frac{I_p\left(2g\sqrt{N_0(N_0+1)}\right)}{[\varepsilon - (\varepsilon_0 - \Delta_{ph}) + i\Gamma/2 + p\cdot\omega_0]} \exp\left(-\frac{p\cdot\omega_0}{2k_BT}\right) \quad (20)$$

where $I_p(x)$ is the $p$th order modified Bessel function and $N_0$ is the Bose distribution function for the optical phonons.

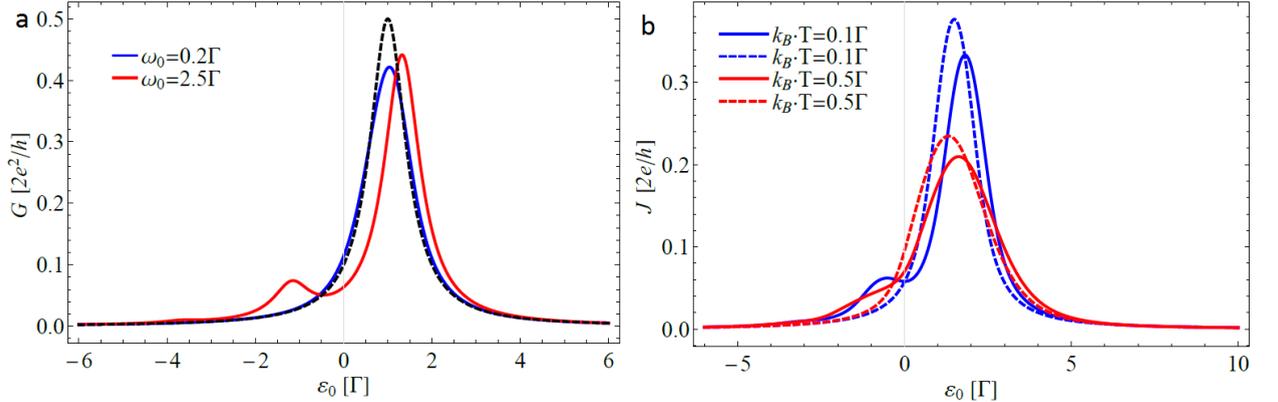

Fig. 2. (a) Conductance as a function of the dot level for different longitudinal optical phonon frequencies at zero temperature. (b) The current as a function of the dot level at finite temperatures, where the phonon frequency is $\omega_0 = 2.5\ \Gamma$ and $eV = 1\ \Gamma$. In both figures the dashed line indicates the case when there is no electron-phonon interaction.

Fig. 2. a and b show the results obtained for the conductance and current as a function of the dot level. First we fixed $\omega_0 = 2.5\ \Gamma$ and $\beta = 0.9\ \Gamma$, then the coupling constant is $g \approx 0.129$ and the energy shift became $\Delta_{ph} \approx 0.324\ \Gamma$. We can see that the peak height is suppressed due to presence of electron-phonon interaction (solid lines) and appears a subpeak. The subpeak refers to the processes when the electrons emit and absorb phonons virtually [7,10]. This curve shows a good similarity with the experimental results [18,28]. It can be seen that the position of the subpeak is modified as a function of the phonon frequency. Thus, increasing the frequency, it is shifted to the negative values of the dot energy. At finite temperatures (Fig. 2 b.) there are real phonon emission and absorption processes [10]. Unlike in the zero temperature case, here the subpeak is suppressed at high values of the temperature.

3.1.2. Interaction with acoustic phonons

Now we consider that the electrons interact with acoustic phonons. In the long-wavelength approximation we obtain:

$$\phi(t-t') \approx i(t-t')\frac{gc}{\sqrt{2\pi}L}\arctan(2\pi) \quad (21)$$

For the energy shift, defined in (9), in the long-wavelength approximation we find:

$$\Delta_{ph} = \sqrt{2\pi}\,gc/L \quad (22)$$

The retarded Green function becomes:

$$G^r(\varepsilon) = \frac{1}{\varepsilon - \varepsilon_0 + \sqrt{2\pi}\left[1 - \frac{\arctan(2\pi)}{2\pi}\right]\frac{gc}{L} + i\Gamma/2} \quad (23)$$

At finite temperatures, we also assume that the wave vector tends to zero. The shift energies remain unchanged. Thus, the final Green functions are equal to the results of zero temperature case and therefore we get the same result for both, zero and non-zero temperature cases for the Green functions.

Fig. 3 a. presents the numerical results for the conductance at zero temperature. For the dimensionless coupling parameter, $g$, we have chosen 0.05 for a GaAs type dot [6-7,10]. The speed of sound is $c = 3\,\Gamma \cdot L$. We can observe that there is no subpeak, which refers to the fact that the spectrum of acoustic phonons is continuous, as expected. Fig. 3 b. shows the currents as a function of dot energy at different temperatures. It is evident that the peak is suppressed at high temperature values.

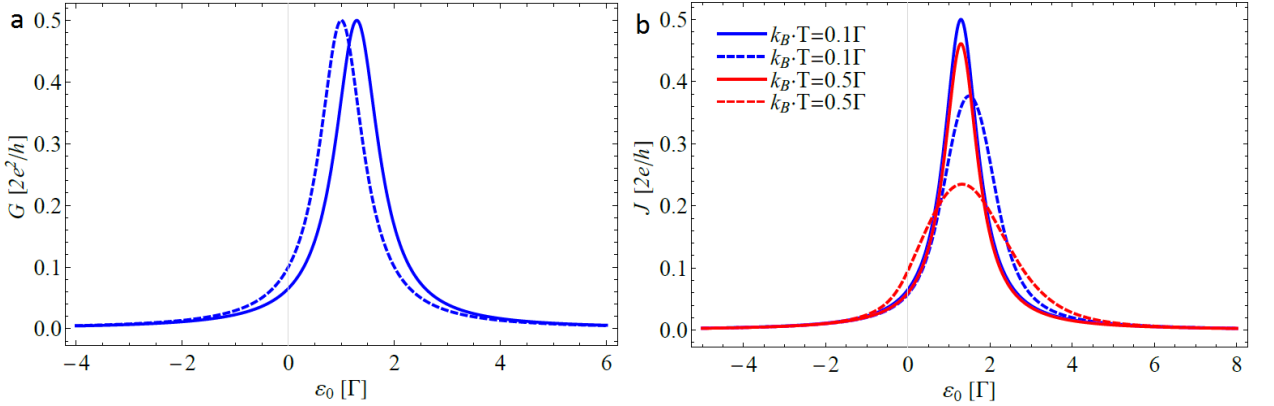

Fig. 3. (a) Conductance as a function of the dot level at zero temperature. (b) Current as a function of dot level at different temperatures for eV = 1 Γ. In both figures the dashed lines indicate the no electron-phonon interaction cases.

*3.2. Time-averaged current*

We use a similar form for time-dependence of the bias voltage made by A.-P. Jauho et al. [17] for different contacts modulations without electron-phonon interaction. The dot energy and the contacts modulations have the time-dependence:

$$\begin{cases} \varepsilon_0(t) = \varepsilon_0 + \Delta_0 \cos(\omega t) \\ \Delta(t) = \Delta \cos(\omega t) \end{cases} \quad (24)$$

where $\omega$ is the external modulation frequency and $\Delta_0$, $\Delta$ are the modulation amplitudes of dot and contacts.

3.2.1. Interaction with longitudinal optical phonons

We consider the electrons interacting with optical phonons, at zero temperature. To determine the averaged conductance and current, we calculate:

$$A(\varepsilon,t) = e^{-g}\, e^{-i\frac{(\Delta_0 - \Delta)}{\omega}\sin(\omega t)} \sum_{n=0}^{\infty} \frac{g^n}{n!} \sum_{r=-\infty}^{\infty} J_r\left(\frac{\Delta_0 - \Delta}{\omega}\right) \frac{\exp(i\,r\omega t)}{\varepsilon - (\varepsilon_0 - \Delta_{ph}) - \omega_0 n - r\cdot\omega + i\Gamma/2} \quad (25)$$

At finite temperature the function $A(\varepsilon,t)$ is:

$$A(\varepsilon,t) = e^{-g(2N_0+1)} e^{-i\frac{(\Delta_0-\Delta)}{\omega}\sin(\omega t)} \sum_{p=-\infty}^{\infty} I_p\left(2g\sqrt{N_0(N_0+1)}\right) \exp\left(-\frac{p \cdot \omega_0}{2k_B T}\right) \times$$
$$\times \sum_{r=-\infty}^{\infty} J_r\left(\frac{\Delta_0-\Delta}{\omega}\right) \frac{\exp(ir\omega t)}{\varepsilon-(\varepsilon_0-\Delta_{ph})+p\cdot\omega_0-r\cdot\omega+i\Gamma/2} \quad (26)$$

where $I_p(x)$ is the $p$th order modified Bessel function and $J_r(x)$ represents the $r$th order Bessel function. The imaginary parts of Eqs. (25) and (26) give the time averaged conductance and current flowing from the left contact to the quantum dot. The results are shown in Fig. 4 and Fig. 5.

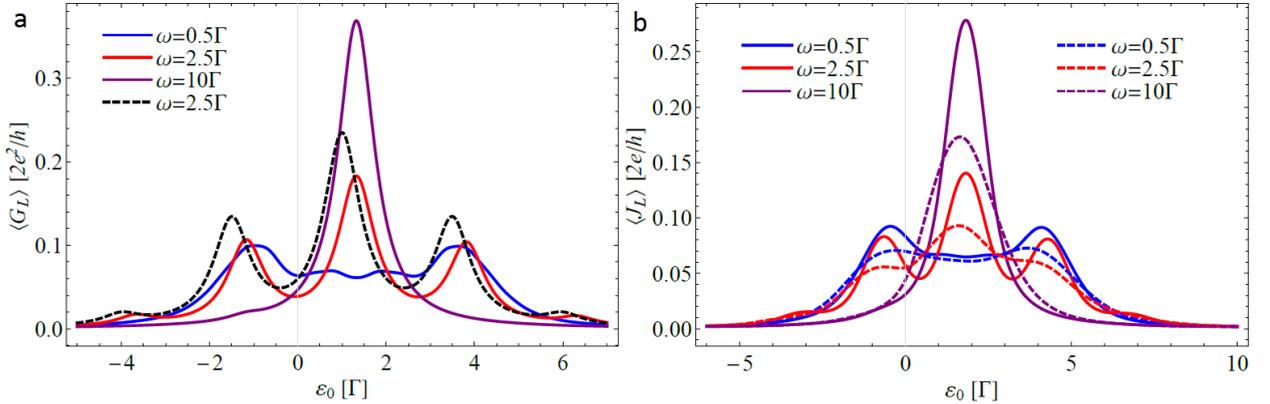

Fig. 4 (a) Conductance from left contact as a function of dot level for different modulation frequencies, when $\omega_0 = 2.5\ \Gamma$. The dashed line refers to the no interaction case. (b) The current as a function of dot level for different values of the modulation frequency, where the solid lines correspond to $k_B T = 0.1\ \Gamma$ and the dashed lines to $k_B T = 0.5\ \Gamma$ with $eV = 1\ \Gamma$.

First, similarly to the stationary case, we fix the phonon frequency at $\omega_0 = 2.5\ \Gamma$ and $\beta = 0.9\ \Gamma$, then the coupling constant becomes $g \approx 0.129$ and the energy shift is $\Delta_{ph} \approx 0.324\ \Gamma$. In Fig. 4 a. the conductance from left contact is plotted as a function of dot level at zero temperature for different values of the modulation frequency of the contacts. It can be seen that, for high modulation frequencies, we get a similar result to the time-independent case, so we can observe the subpeak caused by the phonon mode. When $\omega$ decreases the subpeak disappears and many parasite peaks show up in the spectrum. The Fig. 4 b. shows the current from left contact for different temperatures and in Fig. 5 we plotted the conductance and the current from left contact as a function of dot level for fixed modulation frequency and different phonon modes. We observe that at low phonon frequencies the subpeak disappears and at high temperatures the main maximum is compressed.

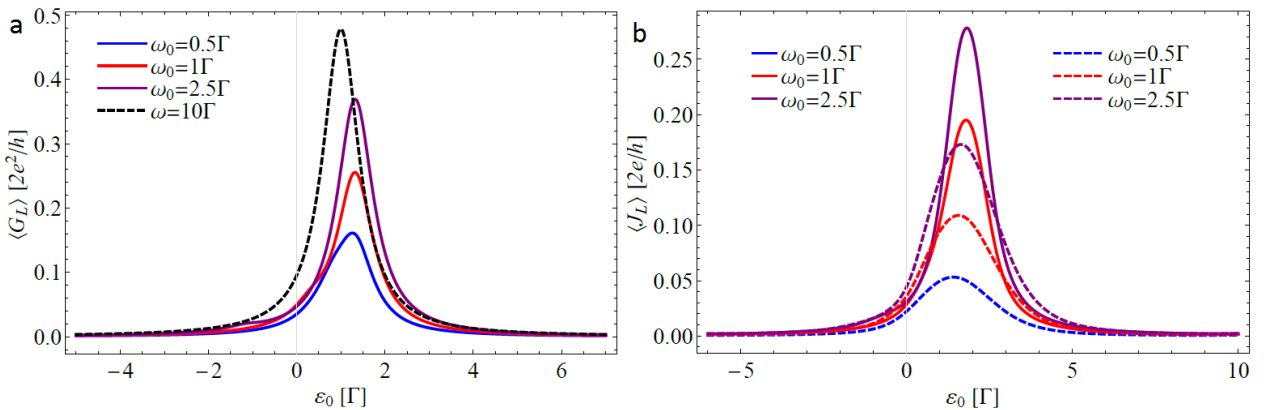

Fig. 5 (a) Conductance as a function of the dot level for different phonon frequency, when the modulation is ω = 10 Γ. The dashed line indicates the case when there is no electron-phonon interaction. (b) Current as a function of dot level for eV = 1 Γ. Solid lines correspond to $k_B T$ = 0.1 Γ and the dashed lines indicate the case of $k_B T$ = 0.5 Γ. In both figures the modulation amplitudes are: $\Delta_0$ = 3 Γ, $\Delta$ = 6 Γ.

3.2.2. Interaction with acoustic phonons

We have seen that in the case of electron-acoustic phonon interaction it is necessary to make an approximation for the wave vectors of phonons. Also, we found that the function $\phi(t-t')$ has the same form at zero and non-zero temperatures. Thus for the function $A(\varepsilon,t)$ we find:

$$A(\varepsilon,t) = e^{-i\frac{(\Delta_0-\Delta)}{\omega}\sin(\omega t)} \sum_{r=-\infty}^{\infty} J_r\left(\frac{\Delta_0-\Delta}{\omega}\right) \frac{\exp(ir\omega t)}{\varepsilon-\varepsilon_0 + \sqrt{2\pi}\left[1-\frac{\arctan(2\pi)}{2\pi}\right]\frac{gc}{L} - r\cdot\omega + i\Gamma/2} \quad (27)$$

Using this relation, we can calculate the conductance and the current for finite temperatures. In Fig. 6 a. the conductance is plotted as a function of dot level at zero temperature for different values of the modulation frequency. The used parameters are shown in the description of figure. The Fig 6 b. shows the current as a function of dot energy for different temperatures. The dashed lines denote the cases when there is no interaction.

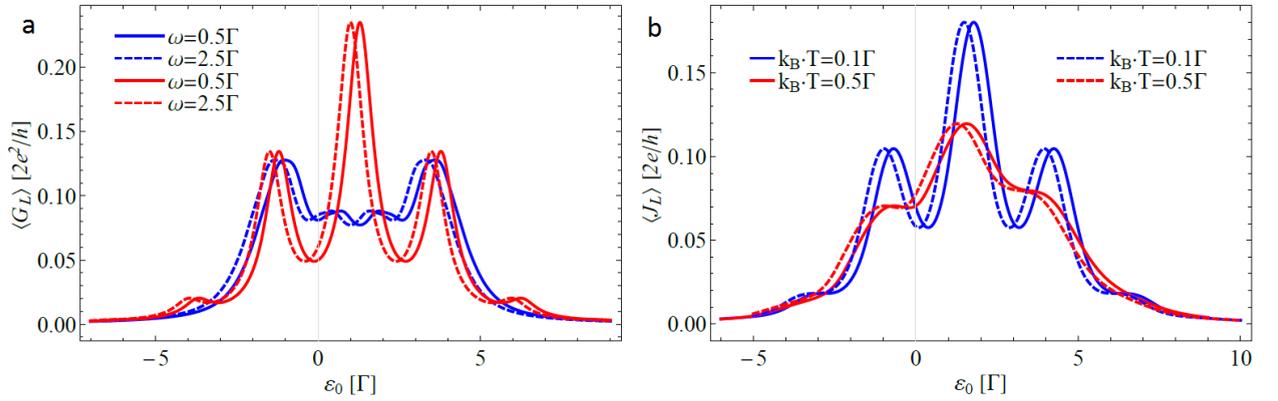

Fig. 6. (a) Conductance from left contact as a function of the dot level for different modulation frequencies at zero temperature. (b) The current from left contact as a function of the dot level for different temperatures when ω = 2.5 Γ with eV = 1 Γ. The dashed lines indicate the no interaction cases. In both figures we have: $\Delta_0$ = 3 Γ and $\Delta$ = 6 Γ.

## 4. Conclusions

In the present work we have investigated the transport properties of a quantum dot coupled to two metallic leads in the presence of electron-phonon interaction. In the case when the dot electrons interact with the longitudinal optical phonons in the spectrum appears a subpeak with a position that depends on the phonon frequency. When the dot electrons interact with the acoustic phonons the spectrum is continuous for all temperatures. Here we have assumed the long wavelength limit for the phonon modes.

Also, we have studied the time-averaged current at zero and finite temperatures of a single level quantum dot coupled to longitudinal optical or acoustic phonon modes. We have established that in all situations many parasite peaks appear in the spectrum, due to the external time modulation. In the case of optical phonons, it is very difficult to find the respective subpeak. We can observe that the transport properties strongly depend on the time-modulation frequency.